\documentclass[11pt,a4paper]{article}
\pdfoutput=1

\usepackage{jheppub}
\usepackage[T1]{fontenc}
\usepackage{lmodern}
\usepackage{mathtools}
\usepackage{microtype}
\usepackage{xcolor}
\usepackage{amsmath,amssymb,amsthm,mathtools}
\usepackage{enumitem}
\usepackage{booktabs,tabularx,array}

\newif\ifcomments \commentstrue   

\newcommand{\tr}{\operatorname{tr}}

\newcommand{\arxiv}[1]{\href{https://arxiv.org/abs/#1}{arXiv:#1}}

\numberwithin{equation}{section}
\allowdisplaybreaks[1]
\title{Fortuitous BMN Operators from Multiple Cores}

\author[a,b]{Robert de Mello Koch,}
\author[c]{Sarthak Duary,}
\author[a,d]{Minkyoo Kim}
\author[e]{and Sinong Liu}

\affiliation[a]{School of Science, Huzhou Normal University, Huzhou 313000, China}
\affiliation[b]{Mandelstam Institute for Theoretical Physics, School of Physics,\\
University of the Witwatersrand, Private Bag 3, Wits 2050, South Africa}
\affiliation[c]{Yau Mathematical Sciences Center (YMSC), Tsinghua University, Beijing 100084, China}
\affiliation[d]{Research Institute of Basic Sciences, Seoul National University, Seoul 08826, Republic of Korea}
\affiliation[e]{Department of Physics, Kyung Hee University, 26, Kyungheedae-ro, Dongdaemun-gu,\\ Seoul, 02447, Republic of Korea}

\emailAdd{robert@zjhu.edu.cn}
\emailAdd{sarthakduary@tsinghua.edu.cn}
\emailAdd{mimkim80@gmail.com}
\emailAdd{sinong.liu@khu.ac.kr}

\abstract{We develop a systematic construction of fortuitous cohomology classes in the BMN sector of $\mathcal N=4$ super Yang--Mills theory from multiple lower-rank building blocks. For arbitrary partitions of colour space, we construct corrected effective letters that obey the supersymmetry transformations of the corresponding lower-rank theories. These allow us to build gauge-invariant, $Q$-closed polynomial operators. A restriction map recovers the constituent cohomology classes, establishing non-exactness and fortuity without brute-force computations. As an explicit example, we combine two fortuitous $SU(2)$ cores to construct a fortuitous class in $SU(4)$. Our results provide a systematic approach to the structure of fortuitous cohomology at finite $N$.}

\begin{document}
\maketitle
\flushbottom

\section{Introduction}\label{sec:intr}

The AdS/CFT correspondence identifies the gravitational coupling with $1/N^2$, making the study of CFT dynamics at finite $N$ an important window into quantum gravity at finite gravitational coupling. The transition from large to finite $N$ also has profound structural consequences for the gauge-invariant operators of the theory, independent of the details of its interactions. At large $N$, gauge-invariant operators are freely generated by single traces. At finite $N$, however, the single traces are no longer independent: they satisfy nonlinear polynomial identities known as trace relations~\cite{dMKJ1,dMKGVZ,dMKKVZ,dMKJ2,CdMK,dMKKMR}. These relations modify both the number of independent generating invariants and the algebraic structure of the space of gauge-invariant operators~\cite{dMKJ1,dMKJ2}.

In supersymmetric gauge theories, trace relations have an even more remarkable consequence: they can change the cohomology of the supercharge $Q$. An operator that is not $Q$-closed at large $N$ can become closed at a particular finite $N$ precisely because its $Q$-variation vanishes as a consequence of a trace relation. The resulting cohomology classes, known as \emph{fortuitous classes}~\cite{ChangLin24,ChangLin22,CKLP,CKLLP}, have no ancestors in the large-$N$ cohomology. Fortuity thus reveals a previously unappreciated mechanism by which genuinely new supersymmetric states can emerge at finite $N$. This possibility is especially significant for holography: fortuitous classes are natural candidates for the microstates of $1/16$-BPS black holes and may help explain how the enormous number of states required to account for their entropy arises in the dual CFT. The phenomenon has been demonstrated explicitly in the BMN sector of $\mathcal N=4$ super Yang--Mills theory~\cite{GGKM,ChangYin}, with concrete fortuitous classes constructed at small rank~\cite{ChangLin22,CKLP,CKLLP,CCKLL,BFops,KLLS26}. Fortuity has since been identified in a variety of other systems~\cite{CCSY,Chen,KLLO,BSVZ,BdG1,GIR,BdG2,ChoiKim,CCK,FJP,Chryssanthacopoulos,Berkooz}, alongside a growing body of related developments~\cite{BMV,HS,GLRT,ChangLin25,ChoiLee,vLMR,ChangZhang,GKvLM,KvLM,KKLP26,FortuityAI}.

With this motivation it is clearly of great interest to find systematic methods for constructing fortuitous operators. The earliest examples were discovered through direct numerical computations of the supercharge cohomology at small $N$~\cite{ChangLin22}. A more constructive approach was subsequently developed using syzygies, or relations among trace relations, to build ans\"atze for fortuitous operators~\cite{CCKLL}. In this approach, $Q$-closure follows directly from the finite-$N$ trace relations. The more difficult problem is establishing that the resulting operator represents a nontrivial cohomology class, rather than being $Q$-exact or cohomologous to a graviton operator. Establishing non-exactness has generally required a brute-force computation of the complete space of $Q$-exact operators, followed by an explicit verification that the candidate operator does not lie in their linear span. The computational complexity of this procedure has been a major obstacle to the systematic construction of fortuitous operators. Nevertheless, explicit constructions have been achieved for $SU(2)$, $SU(3)$ and, more recently, $SU(4)$~\cite{ChangLin22,CKLP,CKLLP,CCKLL,BFops,KLLS26}.

An important advance came with the brane-fused construction, which provided a new way to build fortuitous operators in the $SU(3)$ theory~\cite{BFops}. The central idea is to use the known $SU(2)$ fortuitous operator as a nonabelian core and combine it with additional degrees of freedom to obtain a genuine $SU(3)$ cohomology class. The construction offers a much more efficient way to establish non-exactness: by restricting the $SU(3)$ operator to an $SU(2)$ core, one recovers the original nontrivial cohomology class, thereby proving that the larger operator cannot be $Q$-exact. This idea has since been extended to $SU(4)$, where an $SU(3)$ brane-fused operator serves as the building block \cite{KLLS26}. More recently, the underlying principle has been developed into a general construction valid for every $SU(N)$, $N\geq 3$ \cite{DKL}. By embedding a corrected two-colour core into $N\times N$ matrices and exploiting an exact restriction identity, one can construct gauge-invariant, $Q$-closed operators and establish their non-exactness and fortuity without relying on brute-force cohomology computations. This provides a systematic mechanism for constructing fortuitous cohomology classes at arbitrary rank.

In this paper, we extend this idea by developing a general framework for constructing fortuitous operators from multiple lower-rank building blocks. Consider any partition $N=n_1+n_2+\cdots+n_k$, together with a collection of fortuitous operators, one for each of the corresponding lower-rank theories. We show how to combine these building blocks to construct new multicore operators in the rank-$N$ theory. The key observation is that the original $N\times N$ BMN letters can be used to construct $k$ sets of effective letters, each associated with one of the smaller blocks. Contributions from the off-diagonal matrix elements implies that simply projecting the original matrices onto these blocks does not preserve the supersymmetry transformations. However, these unwanted contributions can be cancelled by suitable corrections, so that each set of effective letters obeys the BMN supersymmetry transformations of the corresponding lower-rank theory. Substituting these effective letters into the known fortuitous operators therefore produces $Q$-closed building blocks. Taking products of the $Q$-closed operators constructed from these blocks, multiplying by suitable polynomials in the eigenvalues to cancel the denominators introduced by the construction, and summing over the possible partitions of the colour indices, we obtain gauge-invariant polynomial operators of the full rank-$N$ theory.

It remains to establish that these operators define genuinely new fortuitous cohomology classes. We address this using a restriction map that separates the different cores and recovers their original lower-rank cohomology classes. By introducing a parameter that controls the separation between the cores, we show that a suitable coefficient in the restricted operator isolates the product of the original building blocks. Since the restriction commutes with the supercharge $Q$, the non-exactness of the constituent classes implies the non-exactness of the resulting multicore operator. Further, by tracking the action of the restriction map on the stable cohomology, we establish that the new class has no large-$N$ ancestor and is therefore fortuitous, provided the detected product lies outside the stable sector. This gives a systematic route to constructing fortuitous operators from known lower-rank examples, replacing the computationally expensive verification of non-exactness with an argument based on the cohomology of the constituent cores.

The paper is organized as follows. In Section~\ref{effletersec}, we construct effective BMN letters associated with arbitrary partitions of the colour space and show that they obey the supersymmetry transformations of the corresponding lower-rank theories. In Section~\ref{ginvsec}, we explain how to remove the denominators introduced by this construction and obtain genuine gauge-invariant polynomial operators in the original matrix letters. In Section~\ref{DetcorSec}, we develop the restriction map that allows us to recover the constituent cohomology classes and establish both non-exactness and fortuity. Section~\ref{sec:ex22} illustrates the construction explicitly by combining two copies of the $SU(2)$ fortuitous operator $O_{24}$ into a fortuitous operator in the $SU(4)$ theory. We discuss the implications of our results and possible directions for future work in Section~\ref{sec:discussion}. An additional example involving the field-strength letter $f$ is presented in Appendix~\ref{Appflett}. 

\section{Effective BMN letters for any partition of colour space}\label{effletersec}

In this section, we construct effective $n_a\times n_a$ BMN letters embedded into the original $N\times N$ matrices, for any partition $N=n_1+\cdots+n_k$. By effective BMN letters, we mean matrices that obey the same BMN supersymmetry transformation algebra as the letters of the rank-$n_a$ theory. More precisely, there exists a basis in which all the effective letters associated with a given core have support on the same $n_a\times n_a$ block, with all matrix elements outside this block vanishing. All that is used in this construction is the original $N\times N$ supersymmetry transformations. 

Denote the traceless $N\times N$ BMN letters by $\phi^m,\psi_m,f$ ($m=1,2,3$). The scalars $\phi^m$ and $f$ are even, and the fermions $\psi_m$ are odd. $Q$ acts as follows
\begin{eqnarray}
&&Q\phi^m=0,\qquad Q\psi_1=[\phi^2,\phi^3],\qquad Q\psi_2=[\phi^3,\phi^1],\cr\cr 
&&Q\psi_3=[\phi^1,\phi^2],\qquad Qf=\sum_{m=1}^3[\phi^m,\psi_m].\label{eq:Qtwo}
\end{eqnarray}
When acting on a product $Q$ obeys the usual graded product rule. Further, we assume that the supercharge obeys the Leibniz rule which needs to be revisited once loop corrections are included. We construct effective letters for each block in the partition introduced above
\begin{equation}
\{1,\ldots,N\}=S_1\sqcup\cdots\sqcup S_k,\qquad |S_a|=n_a,\qquad \sum_{a=1}^k n_a=N.\label{eq:partition}
\end{equation}
The effective letters must obey the algebra \eqref{eq:Qtwo}. For an $SU(n_a)$ core we use the traceless part of each $n_a\times n_a$ block. Blocks of size one have the trivial traceless algebra; they are spectators. Our construction starts by singling out a scalar, say $\phi^1$. The construction is most transparent in the eigenbasis of $\phi^1$. Let $H=\phi^1$. On the open set where $H$ has distinct eigenvalues, we choose a basis with
\begin{equation}
H={\rm diag}(\lambda_1,\ldots,\lambda_N),\qquad \sum_i\lambda_i=0.\label{eq:H}
\end{equation}
Since $QH=0$, the eigenvalues, spectral projectors, and their rational functions are all annihilated by $Q$. For any matrix $Y$,
\begin{equation}
[H,Y]_{ij}=(\lambda_i-\lambda_j)Y_{ij},\label{eq:commH}
\end{equation}
so it is simple to invert the action of the commutator
\begin{equation}
[H,\hat Y]_{ij}=Y_{ij}\qquad\hat{Y}_{ij}= \frac{Y_{ij}}{\lambda_i-\lambda_j}
\end{equation}
This simple answer is why we work in this basis. Fix a block $S=S_a$ and write $R=S^c$ for \emph{all} remaining indices, irrespective of how they are partitioned. For $i,j\in S$, define the effective scalar letter $B^m_{ij}=\phi^m_{ij}$. Since $QB^m_{ij}=0$ we reproduce the first equation in \eqref{eq:Qtwo} and this is a good effective letter. Based on this success we try to similarly restrict $\psi_1$. This does not produce a valid effective letter. Indeed, writing $Q\psi_1=[\phi^2,\phi^3]$, out in detail we find
\begin{align}
Q\,\,(\psi_1)_{ij}&=\sum_{\ell\in S}(\phi^2_{i\ell}\phi^3_{\ell j}-\phi^3_{i\ell}\phi^2_{\ell j})
 +\underbrace{\sum_{\ell\in R}(\phi^2_{i\ell}\phi^3_{\ell j}-\phi^3_{i\ell}\phi^2_{\ell j})}_{E_{ij}}
 =[B^2,B^3]_{ij}+E_{ij}.\label{eq:split}
\end{align}
The $E_{ij}$ term spoils the match to the algebra \eqref{eq:Qtwo}. We need to add a correction to $(\psi_1)_{ij}$ whose $Q$ variation cancels $E_{ij}$. To motivate the form of the correction, note that the other fermion transformations after using \eqref{eq:commH}, give
\begin{equation}
Q(\psi_3)_{i\ell}=(\lambda_i-\lambda_\ell)\phi^2_{i\ell},\qquad
Q(\psi_2)_{i\ell}=-(\lambda_i-\lambda_\ell)\phi^3_{i\ell}.\label{eq:qpsi23}
\end{equation}
When $i\in S$ and $\ell\in R$, the denominator is nonzero. Hence
\begin{equation}
Q\left(\frac{(\psi_3)_{i\ell}}{\lambda_i-\lambda_\ell}\right)=\phi^2_{i\ell},\qquad
Q\left(\frac{(\psi_2)_{i\ell}}{\lambda_i-\lambda_\ell}\right)=-\phi^3_{i\ell}.\label{eq:qquotients}
\end{equation}
Plugging these into $E_{ij}$ it is obvious that $QK_{ij}=E_{ij}$ with
\begin{equation}
K_{ij}=\sum_{\ell\in R}\frac{(\psi_3)_{i\ell}(\phi^3)_{\ell j}+(\psi_2)_{i\ell}(\phi^2)_{\ell j}}{\lambda_i-\lambda_\ell},\qquad i,j\in S.\label{eq:Ktwo}
\end{equation}
Indeed, applying \eqref{eq:qquotients} term by term to the two $\psi_m$'s gives
\begin{equation}
QK_{ij}=\sum_{\ell\in R}(\phi^2_{i\ell}\phi^3_{\ell j}-\phi^3_{i\ell}\phi^2_{\ell j})=E_{ij}.
\label{eq:qK}
\end{equation}
Thus, with $\widehat\psi_1=(\psi_1)_{SS}-K$, we have the correct algebra
\begin{equation}
Q\widehat\psi_1=[B^2,B^3]\label{eq:qPsi1}
\end{equation}
which proves that $\widehat\psi_1$ is a valid effective letters. It is straight forward to verify that $\widehat\psi_2=(\psi_2)_{SS}$ and $\widehat\psi_3=(\psi_3)_{SS}$ also define valid effective letters. Indeed, since $H$ is diagonal, we have
\begin{equation}
Q(\widehat\psi_2)=[B^3,B^1],\qquad Q(\widehat\psi_3)=[B^1,B^2].\label{eq:otherpsi}
\end{equation}
Thus, the three effective fermionic letters and their correct transformations are established.

Up to this point our analysis has been performed in the eigenbasis of $H$. Ultimately we would like to write the construction in a manifestly basis independent language. Towards this end we introduce the diagonal projector $P$, which projects onto $S$, and $R_P={\mathbf 1}-P$ which projects onto its orthogonal complement. In terms of $P$ we have $B^m=P\phi^mP$. Denote $J_m=P\psi_mP$. Define fermionic \emph{off-block} matrices $X_2,X_3$ by
\begin{equation}
(X_m)_{i\ell}=\frac{(\psi_m)_{i\ell}}{\lambda_i-\lambda_\ell}\qquad i\in S,\ \ell\in R,\qquad\qquad\qquad X_m=PX_mR_P.\label{eq:Xdef}
\end{equation}
They obey
\begin{equation}
[H,X_m]=P\psi_mR_P,\qquad QX_3=P\phi^2R_P,\qquad QX_2=-P\phi^3R_P.\label{eq:Xrelations}
\end{equation}
In terms of $X_2$ and $X_3$ we can write \eqref{eq:Ktwo} as
\begin{equation}
K=X_3\phi^3P+X_2\phi^2P.\label{eq:Km}
\end{equation}
We now turn to the construction of the effective letter corresponding to $f$. It is simple to verify that simply projecting $f$ to $f_{ij}$ does not produce a valid effective letter. Towards this end, consider the transformation of $PfP$. Inserting ${\mathbf 1}=P+R_P$ into $P[\phi^m,\psi_m]P$ gives
\begin{align}
Q(PfP)&=P(Qf)P=\sum_{m=1}^3P[\phi^m,\psi_m]P\nonumber\\
&=\sum_{m=1}^3\left(P\phi^m(P+R_P)\psi_mP
-P\psi_m(P+R_P)\phi^mP\right)\nonumber\\
&=\sum_{m=1}^3[B^m,J_m]+E_f,\label{eq:qfblock}
\end{align}
where
\begin{align} 
E_f=\sum_{m=2,3}\big(P\phi^mR_P\psi_mP-P\psi_mR_P\phi^mP\big).
\end{align}
There is no $m=1$ contribution to $E_f$, because $H$ commutes with $P$ so that $P\phi^1 R_P=PH R_P=P R_PH=0$. The corrected field $F=PfP-L$ must obey
\begin{equation}
QF=[B^1,J_1-K]+[B^2,J_2]+[B^3,J_3].\label{eq:desiredF}
\end{equation}
Comparing with \eqref{eq:qfblock} shows that the correction $L$ must obey
\begin{equation}
QL=E_f+[B^1,K].\label{eq:needL}
\end{equation}
The bosonic correction (a product of two fermionic matrices) is given by
\begin{equation}
L=X_3\psi_2P-X_2\psi_3P.\label{eq:L}
\end{equation}
This is easily checked (recall that $X_m$ is fermionic)
\begin{align}
QL&=(QX_3)\psi_2P-X_3(Q\psi_2)P-(QX_2)\psi_3P+X_2(Q\psi_3)P\nonumber\\
&=P\phi^2R_P\psi_2P+P\phi^3R_P\psi_3P+X_3[H,\phi^3]P+X_2[H,\phi^2]P.\label{eq:qL}
\end{align}
Applying $[H,AB]=[H,A]B+A[H,B]$ to \eqref{eq:Km}, and using \eqref{eq:Xrelations}, gives
\begin{align}
[B^1,K]=[H,K]&=P\psi_3R_P\phi^3P+P\psi_2R_P\phi^2P+X_3[H,\phi^3]P+X_2[H,\phi^2]P.\label{eq:HcommK}
\end{align}
Subtracting \eqref{eq:HcommK} from \eqref{eq:qL} gives $E_f$ in \eqref{eq:qfblock} and so $QL=E_f+[B^1,K]$. Substituting into \eqref{eq:qfblock} proves \eqref{eq:desiredF}. In components the field letter correction is
\begin{equation}
L_{ij}=\sum_{\ell\in R}\frac{(\psi_3)_{i\ell}(\psi_2)_{\ell j}-(\psi_2)_{i\ell}(\psi_3)_{\ell j}}{\lambda_i-\lambda_\ell},\qquad i,j\in S.\label{eq:Lcomponents}
\end{equation}
This completes the correction of the effective letters. 

A comment is in order. Working in the eigenbasis of $H=\phi^1$ has made the above argument transparent. In view of gauge invariance, we must in the end write the construction in a manifestly basis-independent language. This can be achieved by using the spectral projector for subset $S$
\begin{equation}
P_S=\sum_{i\in S}\prod_{j\ne i}\frac{H-\lambda_j{\mathbf 1}_N}{\lambda_i-\lambda_j}.\label{eq:projector}
\end{equation}
For a matrix $Y$, the unique off-block solution $X=P_SX({\mathbf 1}-P_S)$ of
\begin{equation}
[H,X]=P_SY({\mathbf 1}-P_S)\label{eq:basisfree}
\end{equation}
is exactly $X_{ij}=Y_{ij}/(\lambda_i-\lambda_j)$ in the eigenbasis. Equations \eqref{eq:projector}--\eqref{eq:basisfree} give the covariant meaning of every correction used above. The construction is rational on the distinct-eigenvalue locus.

\subsection*{Summary: Effective letter  algebra}
We have constructed the effective letters on the subset $S$ of color indices. $S$ refers to indices in the eigenbasis of $H$. Let $n=|S|$. Use the notation
\begin{equation}
Z^0=Z-\frac1n{\rm tr}_n(Z){\mathbf 1}_n\label{DefTrPrt}
\end{equation}
to denote the traceless part of any $n\times n$ matrix $Z$. Subtracting traces commutes with $Q$ because every commutator is traceless and replacing a matrix in a commutator by its traceless part leaves the commutator unchanged. The effective letters are given by
\begin{eqnarray}
&&\widehat A^m=(B^m)^0,\qquad\widehat\psi_1=(J_1-K)^0,\qquad
\widehat\psi_2=(J_2)^0,\cr\cr 
&&\widehat\psi_3=(J_3)^0,\qquad\widehat F=(PfP-L)^0.
\end{eqnarray}
The achievement of the construction given above is that these effective letters obey the BMN algebra. Indeed \eqref{eq:qPsi1}, \eqref{eq:otherpsi}, and \eqref{eq:desiredF} give
\begin{equation}\label{eq:finalalgebra}
Q\widehat A^m=0,\qquad Q\widehat\psi_m=\tfrac12\epsilon_{mnp}[\widehat A^n,\widehat A^p],\qquad
Q\widehat F=\sum_m[\widehat A^m,\widehat\psi_m].
\end{equation}
This is the BMN algebra of $SU(n)$.

Our original partition of $N$ decomposed $N$ into $k$ parts. The above construction immediately provides the effective letters for the blocks defined by each part of the partition. Thus, the original $N\times N$ BMN letters contain $k$ distinct sets of effective letters, one for each block. To extract a particular set, we simply project the original letters onto the chosen block and introduce corrections that restore the $Q$-algebra by cancelling contributions from the remaining colours, exactly as described above. Each set is constructed from the same original $N\times N$ letters. Although we extract only one set at a time, all $k$ sets exist simultaneously and each obeys the BMN $Q$-algebra appropriate to its block size. The construction does not require successive projections or modifications of the original letters.

The different sets of effective letters are not algebraically independent: they involve common off-diagonal entries of the original matrices.  If $c_{n_a}$ is a $Q$-closed trace polynomial in the abstract $SU(n_a)$ BMN letters, we substitute the effective letters for that block and write $C_a=c_{n_a}(\widehat A_a,\widehat\psi_a,\widehat F_a)$. By \eqref{eq:finalalgebra} and the trace identities of the original $n_a$-dimensional theory, $QC_a=0$. The BMN algebra obeyed by the effective letters under the action of $Q$ then proves
\begin{equation}\label{eq:product}
Q(C_1C_2\cdots C_k)=0.
\end{equation}
The factors can be different fortuitous operators and they can have the same rank.

The construction of effective letters does not require the selected blocks to be disjoint; overlapping blocks can also be accommodated. However, to establish non-exactness and fortuity using a restriction map, we need to isolate the individual building blocks and recover their cohomology classes. As we will see, this requires the blocks to be non-overlapping. Finally, we must still show that the operators constructed in this section can be expressed as gauge-invariant polynomials in the original BMN letters, or equivalently, as linear combinations of multi-trace operators. We turn to this problem in the next section.

\section{Obtaining genuine polynomial gauge invariants}\label{ginvsec}

The construction of the previous section produces operators expressed in terms of projections of the BMN letters $\phi^2,\phi^3,\psi_m,f$ and the eigenvalues of $H=\phi^1$. These expressions contain denominators involving differences of eigenvalues, which must be eliminated if the resulting operators are to be written as polynomials in traces of the original BMN letters. We first remove these denominators by multiplying by sufficiently high powers of suitable products of eigenvalue differences. This leaves expressions that are polynomial in the eigenvalues of $H$ and the original BMN letters. The remaining challenge is to eliminate the explicit dependence on individual eigenvalues and express the result entirely in terms of the original matrices. We achieve this by summing over all possible embeddings of the blocks into the $N\times N$ matrices. This produces expressions symmetric in the eigenvalues, allowing their dependence to be rewritten in terms of traces of powers of $H$. The result is a gauge-invariant polynomial in the original BMN letters, providing the desired realization of the construction developed in the previous section. 

Fix the block sizes $\vec n=(n_1,\ldots,n_k)$, with $n_a\geq1$ and $\sum_a n_a=N$. A \emph{labelled partition} is an ordered tuple ${\mathcal P}=(S_1,\ldots,S_k)$ of disjoint subsets of $\{1,\ldots,N\}$ with $|S_a|=n_a$. We refer to a block of size one as a spectator. Since all commutators of $1\times 1$ matrices vanish, the supercharge $Q$ annihilates the corresponding $U(1)$ letters. For the traceless $SU(1)$ theory, these letters vanish identically. Given the trace of a polynomial $c_a$ in the $SU(n_a)$ BMN letters, of degree at most $d_a$, write $C_{S_a}=c_a(\widehat{\mathcal L}_{S_a})$ for the same polynomial evaluated on the effective letters; $C_{S_a}=c_a(\widehat{\mathcal L}_{S_a})$ has denominators given by differences in eigenvalues. For each block $S$, define
\begin{equation}
 \Delta_S=\prod_{i\in S}\prod_{j\notin S}(\lambda_i-\lambda_j),\label{delta}
\end{equation}
where $\lambda_1,\ldots,\lambda_N$ are the eigenvalues (none equal so that $\Delta_S\ne 0$) of $H=\phi^1$. We will argue that for any integer $M_a\geq 5d_a$, all denominators cancel so that the expression
\begin{equation}
 \mathcal O_{\vec n,W}=\sum_{{\mathcal P}=(S_1,\ldots,S_k)} W_{{\mathcal P}}(\lambda_i)
 \prod_{a=1}^k\left[\Delta_{S_a}^{M_a} c_a(\widehat{\mathcal L}_{S_a})\right]\label{master}
\end{equation}
is a polynomial. $W_{\mathcal P}(\lambda_i)$ is a polynomial in the eigenvalues chosen to ensure the appropriate symmetry under exchange of identical cores, while the factors $\Delta_{S_a}^{M_a}$ remove the denominators. We fix an ordering of the core operators in the product. When the cores are distinct, this ordering introduces no ambiguity, and we set $W_{\mathcal P}=1$. When identical fermionic cores are present, for example, exchanging their blocks reverses the sign of the product. An antisymmetric polynomial weight compensates for this sign and prevents cancellation in the sum over partitions. Further, if we are in the end to construct a gauge invariant operator, we must also have $W_{\sigma\mathcal P}(\sigma\cdot\lambda_i)=W_{\mathcal P}(\lambda_i)$. If all $c_a$ are $Q$-closed, then $Q\mathcal O_{\vec n,W}=0$. As we will see below, the bound $5d_a$ is generous; it is not optimal.

We begin by rewriting the projector $P_S$ onto the subspace $S$ in a way that clarifies the structure of the eigenvalue denominator. Define the polynomials
\begin{equation}
q_S(x)=\prod_{i\in S}(x-\lambda_i),\qquad
r_S(x)=\prod_{j\notin S}(x-\lambda_j),
\end{equation}
so that $q_S(x)r_S(x)=\det(xI-H)$. Assuming that the eigenvalues in $S$ are distinct from those in its complement, Bézout's identity~\cite{Bezout} gives
\begin{equation}
a_S(x)q_S(x)+b_S(x)r_S(x)=\Delta_S,\qquad\Delta_S=\prod_{\substack{i\in S\\j\notin S}}(\lambda_i-\lambda_j),
\end{equation}
where $a_S$ and $b_S$ can be chosen with coefficients polynomial in the eigenvalues. Evaluating this identity at $H$, we see that
\begin{equation}
P_S=\frac{b_S(H)r_S(H)}{\Delta_S}
\end{equation}
is the projector onto $S$. Thus, $\Delta_S P_S=b_S(H)r_S(H)$ is polynomial in $H$. Consequently, the denominators appearing in the projector involve only differences between eigenvalues belonging to $S$ and those belonging to its complement.

Apart from the projectors, another ingredient in the corrections to letters is the inverse commutator. We will now show that the denominators appearing in the inverse commutator can also be removed by multiplying by a suitable power of $\Delta_S$. Recall that, in an eigenbasis of $H$,
\begin{equation}
[H,X]_{ij}=(\lambda_i-\lambda_j)X_{ij}.
\end{equation}
Inverting this relation for matrix elements connecting $S$ to its complement requires division by $\lambda_i-\lambda_j$, with $i\in S$ and $j\notin S$. To express this inverse in polynomial form, introduce
\begin{equation}
p_S(z)=\prod_{\substack{i\in S\\j\notin S}}\bigl(z-(\lambda_i-\lambda_j)\bigr),
\qquad p_S(0)=(-1)^{n(N-n)}\Delta_S.
\end{equation}
We can write $p_S(z)=p_S(0)+zh_S(z)$, where $h_S(z)$ is a polynomial. Since $p_S(\lambda_i-\lambda_j)=0$ for every $i\in S$ and $j\notin S$, it follows that
\begin{equation}
\frac{1}{\lambda_i-\lambda_j}=-\frac{h_S(\lambda_i-\lambda_j)}{p_S(0)}.
\end{equation}
This identity allows us to replace division by eigenvalue differences with the action of a polynomial in the commutator with $H$. Towards this end, define ${\rm ad}_H(Y)=[H,Y]$. The inverse commutator can now be written in the basis-independent form
\begin{equation}
X_S(Y)=-\frac{1}{p_S(0)}P_S\,h_S({\rm ad}_H)(Y)\,R_S.
\end{equation}
As a simple check, in the eigenbasis of $H$ this expression reduces to
\begin{equation}
(X_S(Y))_{ij}=\frac{Y_{ij}}{\lambda_i-\lambda_j},\qquad i\in S,\quad j\notin S,
\end{equation}
and vanishes for all other matrix elements. Consequently,
\begin{equation}
[H,X_S(Y)]=P_SYR_S.
\end{equation}
We can now determine the denominators appearing in $X_S(Y)$. Using $\Delta_SP_S=U_S$ and $\Delta_SR_S=\Delta_S I-U_S$, together with $p_S(0)=(-1)^{n(N-n)}\Delta_S$, we obtain
\begin{equation}
\Delta_S^3X_S(Y)
=(-1)^{n(N-n)+1}U_S\,h_S({\rm ad}_H)(Y)(\Delta_S I-U_S).
\end{equation}
Every factor on the right-hand side is polynomial in the original matrix entries and the eigenvalues of $H$. Thus, multiplying the inverse commutator by $\Delta_S^3$ removes its denominators. Thus, just as for the spectral projector, the denominators introduced by the inverse commutator are eliminated using powers of products of eigenvalue differences between $S$ and its complement.

We have now identified the two sources of denominators in our construction: the spectral projectors $P_S$ and the inverse commutators $X_S$. We have shown that multiplying these expressions by $\Delta_S$ and $\Delta_S^3$, respectively, removes their denominators. We can now use these results to determine how many powers of $\Delta_S$ are needed to remove all denominators from the corrected effective letters. This will also give us a sufficient power of $\Delta_S$ to make any operator constructed from these letters polynomial.

Recall that the effective letters are constructed from
\begin{equation}\label{letters}
 B^m=P\phi^mP,\qquad J_m=P\psi_mP,\qquad
 K=X_3\phi^3P+X_2\phi^2P,\qquad
 L=X_3\psi_2P-X_2\psi_3P,
\end{equation}
where $P=P_S$ and $X_m=X_S(\psi_m)$. To obtain the effective $SU(n)$ letters, take the traceless part of each matrix on the subspace selected by $P$. The effective letters are therefore
\begin{equation}
 \widehat A^m=(B^m)^0,\qquad \widehat\psi_1=(J_1-K)^0,\qquad
 \widehat\psi_{2,3}=J_{2,3}^0,\qquad \widehat F=(PfP-L)^0.\label{hats}
\end{equation}

We can now count the powers of $\Delta_S$ needed to remove the denominators. Each projected term $B^m$, $J_m$ or $PfP$ contains two factors of $P$. Since $\Delta_S P$ is polynomial, multiplying these terms by $\Delta_S^2$ is sufficient to remove their denominators. The correction terms $K$ and $L$ each contain one inverse commutator $X_m$ and one additional projector $P$. Since $\Delta_S^3 X_m$ and $\Delta_S P$ are polynomial, multiplying $K$ or $L$ by $\Delta_S^4$ is sufficient.

Note that taking the traceless part, as in \eqref{DefTrPrt}, introduces an additional factor of $P$ through the term proportional to ${\rm tr}_N Z$. This can increase the required power of $\Delta_S$ by one. Consequently, multiplying any of the corrected effective letters by $\Delta_S^5$ is sufficient to remove all denominators. In other words,
\begin{equation}
\Delta_S^5\widehat A^m,\qquad\Delta_S^5\widehat\psi_m,\qquad\Delta_S^5\widehat F
\end{equation}
are all polynomial in the original BMN letters and the eigenvalues of $H$.

This immediately extends from individual letters to operators constructed from them. Consider a polynomial $c_a(\widehat{\mathcal L}_S)$ in the effective letters, and let $d_a$ be the largest number of letters appearing in any monomial of $c_a$. Since each effective letter requires at most five powers of $\Delta_S$, a product of $\ell$ effective letters requires at most $5\ell$ powers. It follows that
\begin{equation}
\Delta_S^{5d_a}c_a(\widehat{\mathcal L}_S)
\end{equation}
is polynomial in the original BMN letters and the eigenvalues of $H$. This gives a simple sufficient condition on the exponents appearing in our construction $M_a\geq 5d_a$. This bound is not optimal, but it guarantees that all denominators are removed. The remaining task is to eliminate the explicit dependence on individual eigenvalues by summing over the possible embeddings of the blocks.

This shows that by choosing the $M_a$ sufficiently large, all denominators in our construction are removed. The resulting expression is polynomial in the original BMN matrices and the eigenvalues $\lambda_1,\ldots,\lambda_N$ of $H$. Our goal is to construct an operator that is polynomial \emph{in the original BMN letters alone}. We must therefore eliminate the explicit dependence on the individual eigenvalues.

Our strategy is to sum over all possible embeddings of the blocks into the $N\times N$ matrices. This produces an expression that is symmetric under permutations of the eigenvalue labels. Standard results from the theory of symmetric polynomials then express this dependence in terms of traces of powers of $H$, eliminating the individual eigenvalues.

There is a subtlety that must be addressed before applying this argument. In the eigenbasis of $H$, the eigenvalues are tied to the matrix indices. For example,
\begin{equation}
(HY)_{ij}=\lambda_iY_{ij},\qquad
(YH)_{ij}=\lambda_jY_{ij},
\end{equation}
so that
\begin{equation}
[H,Y]_{ij}=(\lambda_i-\lambda_j)Y_{ij}.
\end{equation}
We cannot simply permute the eigenvalues while keeping matrix elements fixed. The association between eigenvalues and matrix indices must be preserved. However, this association is already encoded by matrix multiplication. For example,
\begin{equation}
\lambda_i^a\lambda_j^bY_{ij}=(H^aYH^b)_{ij},\qquad {\rm and}\qquad
(\lambda_i-\lambda_j)^2Y_{ij}=\bigl([H,[H,Y]]\bigr)_{ij}.
\end{equation}
Thus, expressions involving eigenvalues attached to matrix indices can be rewritten as polynomial operations on the original matrices. This is the reason for introducing the polynomial $h_S({\rm ad}_H)$ in our construction of the inverse commutator. The dependence on the matrix indices is not discarded; it is incorporated into the matrix algebra.

After canceling denominators, the contribution of a particular partition ${\mathcal P}=(S_1,\ldots,S_k)$ takes the form
\begin{equation}
T_{\mathcal P}=W_{\mathcal P}(\lambda_i)\prod_{a=1}^k
\left[\Delta_{S_a}^{M_a}c_a(\widehat{\mathcal L}_{S_a})\right].
\end{equation}
As just discussed, this can be expressed using polynomial operations on the original matrices, with the remaining dependence on the labelled eigenvalues appearing only in polynomial coefficients. The construction respects relabelling of these eigenvalues. Under a permutation $\sigma\in S_N$, the spectral subset $S_a$ is replaced by $\sigma(S_a)$, and the corresponding contribution is transformed into the contribution associated with the relabelled partition. Provided the weights $W_{\mathcal P}$ transform appropriately, the sum over all partitions is therefore unchanged by permutations of the eigenvalue labels. Consequently,
\begin{equation}
\mathcal O_{\vec n,W}=\sum_{\mathcal P}T_{\mathcal P}
\end{equation}
has coefficients that are symmetric polynomials in $\lambda_1,\ldots,\lambda_N$.

To see how this works in a simple example, consider $N=2$ and the expression
\begin{equation}
F=(H-\lambda_1 I)Y(H-\lambda_2 I)
 +(H-\lambda_2 I)Y(H-\lambda_1 I).
\end{equation}
Each term depends explicitly on the individual eigenvalues, but their sum is symmetric under $\lambda_1\leftrightarrow\lambda_2$. Expanding gives
\begin{equation}
F=2HYH-(\lambda_1+\lambda_2)(HY+YH)
+2\lambda_1\lambda_2Y.
\end{equation}
The eigenvalues now appear only through symmetric combinations. Using
\begin{equation}
\lambda_1+\lambda_2={\rm tr}(H),\qquad
2\lambda_1\lambda_2=({\rm tr}H)^2-{\rm tr}(H^2),
\end{equation}
we obtain
\begin{equation}
F=2HYH-{\rm tr}(H)(HY+YH)
+\left[({\rm tr}H)^2-{\rm tr}(H^2)\right]Y.
\end{equation}
Individual eigenvalues have disappeared, and the result is polynomial in the original matrices. Noncommutative matrix products are preserved throughout this calculation. Only the remaining dependence on the individual eigenvalues is eliminated.

The same argument applies to our general construction. By the fundamental theorem of symmetric polynomials~\cite{Macdonald}, every symmetric polynomial in $\lambda_1,\ldots,\lambda_N$ can be expressed as a polynomial in the elementary symmetric polynomials $e_1,\ldots,e_N$. Newton's identities, in turn, express these in terms of the power sums
\begin{equation}
p_k=\sum_{i=1}^N\lambda_i^k={\rm tr}(H^k),
\qquad k=1,\ldots,N.
\end{equation}
It follows that explicit dependence on individual eigenvalues is eliminated entirely, leaving an expression polynomial in the original BMN matrices and traces of powers of $H$.

We now verify that the resulting operator is gauge invariant and $Q$-closed. The projectors and inverse commutators are constructed using matrix operations that transform covariantly under simultaneous conjugation, so the corrected effective letters also transform by conjugation. The core operators, which are traces of products of these letters, are therefore gauge invariant. Thus, the complete construction is gauge invariant.

Finally, $QH=0$ implies that $Q$ annihilates the eigenvalues, the factors $\Delta_{S_a}$, and the weights $W_{\mathcal P}$. Each set of corrected effective letters obeys the BMN supersymmetry transformations of its corresponding lower-rank theory. Therefore, if the original building blocks satisfy $Qc_a=0$, their products also satisfy $Q=0$, and hence
\begin{equation}
Q\mathcal O_{\vec n,W}=0.
\end{equation}
We have thus established that the multicore construction produces gauge-invariant, $Q$-closed polynomial operators in the original rank-$N$ BMN letters. The remaining question is whether these operators represent nontrivial fortuitous cohomology classes, which we address in the next section.

There is one subtlety worth illustrating. Consider the construction of an $SU(4)$ operator from two identical fermionic $SU(2)$ cores. For a partition into two blocks $S$ and $T$, let $C_S$ and $C_T$ denote the corresponding $Q$-closed operators constructed from the effective letters. Since the cores are fermionic, exchanging the blocks gives
\begin{equation}
C_TC_S=-C_SC_T.
\end{equation}
Consequently, an unweighted sum over ordered partitions vanishes, since the contributions from $(S,T)$ and $(T,S)$ cancel. We can avoid this cancellation by introducing an antisymmetric polynomial weight, for example
\begin{equation}
W_{S,T}=\frac{1}{2}\sum_{i\in S}\lambda_i-\frac{1}{2}\sum_{j\in T}\lambda_j,
\end{equation}
which satisfies $W_{T,S}=-W_{S,T}$. The change of sign in the weight compensates for the change of sign in the product of the fermionic cores. The resulting sum no longer vanishes simply because of the exchange symmetry. This illustrates the role of the weights $W_{\mathcal P}$ in the construction. We return to this example in Section~\ref{sec:ex22}, where we construct an explicit operator using two copies of the $SU(2)$ fortuitous core $O_{24}$.

The construction extends directly to an arbitrary partition $N=n_1+\cdots+n_k$. For each block $S_a$, its complement consists of all the remaining colour indices, irrespective of how these are distributed among the other blocks. The spectral projectors and inverse commutators are therefore constructed independently for each block, using exactly the same procedure developed above. The resulting effective letters may depend on common off-diagonal matrix elements, but this does not affect their supersymmetry transformation properties. After multiplying the corresponding closed core operators, clearing their denominators, and summing over partitions with suitable weights, we obtain a gauge-invariant, $Q$-closed polynomial in the original $SU(N)$ BMN letters.

We emphasize that this construction establishes polynomiality, gauge invariance and $Q$-closure, but does not yet guarantee that the resulting operator is nonzero or represents a nontrivial cohomology class. Even a nontrivial class need not be fortuitous. We address these questions in the next section by developing a restriction map that detects the cohomology of the constituent cores.

\section{Detecting Multicore Cohomology}\label{DetcorSec}

The construction developed above produces gauge-invariant, $Q$-closed polynomial operators in the $SU(N)$ BMN theory. We must still establish that they are not $Q$-exact. We will now develop a method for doing this that avoids the brute-force construction of the space of $Q$-exact operators. The idea is to recover the original lower-rank building blocks from the multicore operator by restricting the $N\times N$ matrices to block-diagonal configurations. We use a parameter $t$ that separates eigenvalues belonging to different blocks. The restricted operator is a polynomial in $t$, and the coefficient of a particular power of $t$ recovers the product of the original building blocks. This restriction commutes with $Q$. Thus, if the recovered product represents a nontrivial cohomology class, the original multicore operator cannot be $Q$-exact.

To make this precise, let ${\mathcal A}_N$ denote the algebra of gauge-invariant polynomials in the traceless $N\times N$ BMN letters. For a partition $N=n_1+\cdots+n_k$, consider the restriction
\begin{equation}
\rho_t(H)=\bigoplus_{a=1}^k(H_a+\alpha_a t I_{n_a}),
\qquad
\rho_t(Y)=\bigoplus_{a=1}^kY_a,
\end{equation}
where $H=\phi^1$, $Y$ denotes any other BMN letter, and $H_a,Y_a$ are traceless. The constants $\alpha_a$ are distinct and $\sum_a n_a\alpha_a=0$, ensuring that the restricted matrix $H$ is traceless. The parameter $t$ controls the separation between eigenvalues of distinct blocks. The BMN supersymmetry transformations involve commutators, and commutators of block-diagonal matrices act independently within each block. Moreover, the shifts $\alpha_a tI_{n_a}$ commute with every matrix. It follows that
\begin{equation}
\rho_t(QF)=Q\rho_t(F),
\end{equation}
where the supercharge on the right acts independently on the lower-rank blocks. Thus, the restriction commutes with $Q$. We make use of this property by extracting a specific coefficient in the expansion in $t$. Define
\begin{equation}
\beta_h(F)=[t^h]\rho_t(F).
\end{equation}
Since $Qt=0$, we immediately have
\begin{equation}
\beta_h(QF)=Q\beta_h(F).
\end{equation}
This gives a test for non-exactness. Suppose ${\mathcal O}$ is $Q$-closed and $\beta_h({\mathcal O})$ represents a nonzero cohomology class in the product of lower-rank theories. If ${\mathcal O}=QB$ in the original rank-$N$ theory, then
\begin{equation}
\beta_h({\mathcal O})=Q\beta_h(B),
\end{equation}
which would make the detected class exact. This is a contradiction. Thus, finding a nontrivial class after restriction establishes that the original multicore operator is not $Q$-exact.

It is useful to illustrate how this works in a concrete example, combining an $SU(2)$ core with an $SU(3)$ core to construct an operator in the $SU(5)$ theory. Let $c_2\in{\mathcal A}_2$ and $c_3\in{\mathcal A}_3$ be homogeneous, $Q$-closed polynomials representing nonzero cohomology classes. For every subset $S\subset\{1,\ldots,5\}$ containing two indices, let $T=S^c$ denote its three-element complement. Recall that
\begin{equation}
\Delta_S=\prod_{i\in S}\prod_{j\notin S}(\lambda_i-\lambda_j),\qquad\Delta_T=\Delta_S.\label{delta}
\end{equation}
Using the corrected effective letters constructed earlier, define
\begin{equation}
{\mathcal O}_{2,3}^{(K)}=\sum_{|S|=2}\Delta_S^K\Delta_T^K\,
c_2(\widehat{\mathcal L}_S)c_3(\widehat{\mathcal L}_T).\label{twocore}
\end{equation}
As established in the previous section, choosing $K\geq\max\{5d_2,5d_3\}$, where $d_a$ is the maximal number of effective letters appearing in a monomial of $c_a$, guarantees that ${\mathcal O}_{2,3}^{(K)}$ is a gauge-invariant polynomial. Its $Q$-closure follows from that of the two building blocks.

We want to show that this operator contains the original product $c_2\otimes c_3$ in a way that can be isolated by our restriction map. To do this, separate the five colours into a physical block $B_2=\{1,2\}$ and a physical block $B_3=\{3,4,5\}$, and take
\begin{equation}
\rho_t(H)=(H_2+3tI_2)\oplus(H_3-2tI_3),\qquad
\rho_t(Y)=Y_2\oplus Y_3.\label{2+3restrict}
\end{equation}
The eigenvalues of the two blocks are then
\begin{equation}
\lambda_i=\mu_i+3t,\quad i\in B_2,\qquad
\lambda_\alpha=\nu_\alpha-2t,\quad \alpha\in B_3.
\end{equation}
Here $\mu_i$ and $\nu_\alpha$ are the eigenvalues of $H_2$ and $H_3$, respectively. The sum over spectral subsets in \eqref{twocore} includes every possible choice of $S$, not only the choice that agrees with the physical blocks.

Consider first the term for which $S=B_2$ and $T=B_3$. The selected blocks coincide with the physical blocks, so all matrix elements connecting the two blocks vanish. The inverse-commutator corrections thus vanish, while the traceless projections remove the shifts proportional to $t$. The effective letters reduce exactly to the original letters of the two lower-rank theories, giving
\begin{equation}
\rho_t\left(c_2(\widehat{\mathcal L}_{B_2})c_3(\widehat{\mathcal L}_{B_3})
\right)=c_2(\mathcal L_2)c_3(\mathcal L_3).\label{recover}
\end{equation}
Further,
\begin{equation}
\rho_t(\Delta_{B_2})=\prod_{i=1}^2\prod_{\alpha=1}^3(5t+\mu_i-\nu_\alpha)
=5^6t^6+O(t^5).\label{six}
\end{equation}
Since $\Delta_{B_3}=\Delta_{B_2}$, the contribution from this term has leading behaviour
\begin{equation}
5^{12K}t^{12K}c_2\otimes c_3.
\end{equation}

If this were the only term contributing at order $t^{12K}$, we would have the result we need. However, the sum in \eqref{twocore} contains nine other choices of $S$. We must show that none of these terms contribute to the same coefficient.

The factors $\Delta_S$ ensure that this is the case. Each eigenvalue difference $\lambda_i-\lambda_j$ is linear in $t$ when the indices belong to different physical blocks. When they belong to the same physical block, the difference is independent of $t$. Thus, the power of $t$ contributed by $\Delta_S$ depends on how the selected subset $S$ overlaps with the physical blocks. Let $x=|S\cap B_2|$. Since $S$ contains two indices, $x$ can take the values $0,1,2$. The number of eigenvalue differences in $\Delta_S$ connecting distinct physical blocks is
\begin{equation}
d(x)=x(x+1)+(2-x)^2.\label{dx}
\end{equation}
The first term counts pairs with the selected index in $B_2$ and unselected index in $B_3$, while the second counts pairs with selected index in $B_3$ and unselected index in $B_2$. We find
\begin{center}
\begin{tabular}{@{}cccl@{}}
\toprule
$x$ & $\deg_t\Delta_S$ & $\deg_t(\Delta_S\Delta_T)$ & Assignment\\
\midrule
$2$ & $6$ & $12$ & Correct\\
$1$ & $3$ & $6$ & Mixed\\
$0$ & $4$ & $8$ & Incorrect\\
\bottomrule
\end{tabular}
\end{center}
The physical block assignment has the highest power $t^{12}$, while every other term has at most $t^8$ in $\Delta_S\Delta_T$. Raising these factors to the power $K$ produces a gap of at least $4K$ powers of $t$ between the physical block assignment and any other term. There is a remaining issue. The corrected effective letters appearing in the unwanted terms may depend on $t$. Can this additional dependence compensate for the smaller power coming from the factors $\Delta_S^K\Delta_T^K$?

The dependence of the effective letters on $t$ is easily determined. On the block-diagonal configurations used in our restriction, spectral projectors are independent of $t$. The inverse commutators also remain bounded as $t$ becomes large. Recall that their matrix elements take the form
\begin{equation}
\frac{Y_{ij}}{\lambda_i-\lambda_j}.
\end{equation}
If $i$ and $j$ belong to different physical blocks, the numerator vanishes because $Y$ is block diagonal. If they belong to the same physical block, the eigenvalue difference is independent of $t$. Thus, inverse commutators are $O(1)$, as are the corrections to the effective fermions and field-strength letter. The only effective letter that can grow with $t$ is $\widehat A^1$, since it is constructed from projections of $H$, whose eigenvalues contain the shifts proportional to $t$. Its growth is at most linear in $t$. Consequently, if $A(c)$ denotes the largest number of occurrences of $A^1$ in any monomial of a core polynomial $c$, then
\begin{equation}\label{growthEq}
c(\widehat{\mathcal L}_S)=O(t^{A(c)}).
\end{equation}
Combining this bound with the powers of $t$ obtained from the eigenvalue differences, we find that every unwanted term grows at most as
\begin{equation}
t^{8K+A(c_2)+A(c_3)}.
\end{equation}
The correct term, by contrast, contains a contribution proportional to $t^{12K}c_2\otimes c_3$, with no additional $t$-dependence from the core operators. Choosing $K$ so that
\begin{equation}
4K>A(c_2)+A(c_3)
\end{equation}
guarantees that no unwanted terms contributes to the coefficient of $t^{12K}$. We thus have the exact identity
\begin{equation}
[t^{12K}]\rho_t\left({\mathcal O}_{2,3}^{(K)}\right)
=5^{12K}c_2\otimes c_3,\label{detectoridentity}
\end{equation}
provided
\begin{equation}
K\geq\max\{5d_2,5d_3\},\qquad 4K>A(c_2)+A(c_3).\label{Kbound}
\end{equation}
The first condition guarantees the operator is polynomial, while the second ensures the restriction isolates the desired product of cores.

This identity is the central result of the example. The K\"unneth theorem gives
\begin{equation}
H_Q({\mathcal A}_2\otimes{\mathcal A}_3)\cong H_Q({\mathcal A}_2)\otimes H_Q({\mathcal A}_3).
\end{equation}
Since $[c_2]$ and $[c_3]$ are nonzero, so   is $[c_2]\otimes[c_3]$. The restriction commutes with $Q$, and therefore the nontriviality of the detected product proves that $[{\mathcal O}_{2,3}^{(K)}]\neq 0$. We have established the non-exactness of the $SU(5)$ multicore operator using the known non-exactness of its $SU(2)$ and $SU(3)$ building blocks.

We have seen how the restriction map establishes non-exactness for an operator constructed from an $SU(2)$ and an $SU(3)$ core. We now consider an important variation: combining two identical fermionic $SU(2)$ cores to construct an operator in the $SU(4)$ theory. This example introduces a new subtlety that is relevant to the general construction.

Let $c_2$ be an odd, $Q$-closed polynomial representing a nonzero cohomology class in the $SU(2)$ theory. Divide the four colours into two sets $S$ and $T$, each containing two indices. Since the cores are fermionic, exchanging blocks changes the sign of their product
\begin{equation}
c_2(\widehat{\mathcal L}_T)c_2(\widehat{\mathcal L}_S)
=-c_2(\widehat{\mathcal L}_S)c_2(\widehat{\mathcal L}_T).
\end{equation}
Furthermore, $\Delta_T=\Delta_S$, so an unweighted sum over ordered partitions vanishes, because the contributions from $(S,T)$ and $(T,S)$ cancel. We avoid this cancellation by introducing the antisymmetric polynomial weight
\begin{equation}
w_{S,T}=\frac12\sum_{i\in S}\lambda_i-\frac12\sum_{j\in T}\lambda_j,\label{weight}
\end{equation}
which measures the difference between the mean eigenvalues of the two blocks. Under an exchange of blocks, $w_{T,S}=-w_{S,T}$, which compensates for the sign change of the fermionic operators. We therefore define
\begin{equation}
{\mathcal O}_{2,2}^{(K)}=\sum_{\substack{S\sqcup T=\{1,2,3,4\}\\|S|=|T|=2}}w_{S,T}
\Delta_S^K\Delta_T^K\,c_2(\widehat{\mathcal L}_S)c_2(\widehat{\mathcal L}_T).\label{2+2op}
\end{equation}
The orderings of each partition now add rather than cancel. As established above, taking $K$ sufficiently large ensures this is a gauge-invariant, $Q$-closed polynomial. We now establish that it represents a nonzero cohomology class. Proceed as in the previous example, restricting the matrices to two physical blocks $B_1$ and $B_2$, each of size two. Specifically, take
\begin{equation}
\rho_t(H)=(H_1+2tI_2)\oplus(H_2-2tI_2),
\end{equation}
with all other BMN letters block diagonal. For the physical block assignment $(S,T)=(B_1,B_2)$, the corrected effective letters reduce to the original $SU(2)$ letters, exactly as before. The factors $\Delta_S\Delta_T$ have leading behaviour $4^8t^8$, while the weight satisfies $\rho_t(w_{B_1,B_2})=4t$. Thus, the contribution from this assignment has leading behaviour
\begin{equation}
4^{8K+1}t^{8K+1}\,c_2\otimes c_2.
\end{equation}
There is now a second physical block assignment, $(S,T)=(B_2,B_1)$. Exchanging the blocks changes the sign of both the weight and the fermionic product, so this assignment makes an identical contribution. The two physical block assignments therefore add.

As before, we must check that the remaining assignments do not contribute at the same order in $t$. There are four such assignments and each divides indices of both physical blocks between $S$ and $T$. For these mixed assignments, $\Delta_S\Delta_T$ grows at most as $t^4$, rather than $t^8$. The leading shifts in the two spectral block means cancel, so $w_{S,T}=O(1)$.

The dependence of the corrected core operators on $t$ is bounded, exactly as in the $SU(2)\times SU(3)$ example. Each core grows at most as $t^{A(c_2)}$. Consequently, the mixed assignments grow at most as $t^{4K+2A(c_2)}$. Choosing $4K>2A(c_2)$ ensures that the mixed assignments cannot contribute to the coefficient of $t^{8K+1}$. Together with the polynomiality bound on $K$, this gives the exact identity
\begin{equation}
[t^{8K+1}]\rho_t({\mathcal O}_{2,2}^{(K)})=8\cdot 4^{8K}\,c_2\otimes c_2.\label{22identity}
\end{equation}
The numerical factor $8$ comes from the two physical block assignments, each contributing a factor of $4$ from the weight. Since $[c_2]\neq0$, the K\"unneth theorem implies that $[c_2]\otimes[c_2]$ is nonzero. The restriction map therefore proves that ${\mathcal O}_{2,2}^{(K)}$ is not $Q$-exact.

These two examples illustrate the general mechanism. The factors $\Delta_{S_a}$ favor spectral partitions that agree with the physical blocks, while suitable weights prevent cancellations between equivalent assignments. We now argue that the same mechanism works for an arbitrary partition $N=n_1+\cdots+n_k$.

Let the physical blocks be $B_1,\ldots,B_k$, with $|B_b|=n_b$, and introduce distinct shifts $\alpha_b t$ as in our general restriction map. Each term in the multicore operator is associated with a spectral partition ${\mathcal P}=(S_1,\ldots,S_k)$, where $|S_a|=n_a$. To compare a spectral partition with the physical blocks, define
\begin{equation}
m_{ab}=|S_a\cap B_b|.\label{matrix}
\end{equation}
$m_{ab}$ counts how many indices belonging to the physical block $B_b$ are assigned to the spectral block $S_a$. Since the spectral and physical partitions both contain all $N$ indices,
\begin{equation}
\sum_bm_{ab}=n_a,\qquad \sum_am_{ab}=n_b.
\end{equation}
The entries of this intersection matrix determine the $t$-dependence of the factors $\Delta_{S_a}$. Recall that
\begin{equation}
\Delta_{S_a}=\prod_{\substack{i\in S_a\\j\notin S_a}}(\lambda_i-\lambda_j).
\end{equation}
A factor $\lambda_i-\lambda_j$ grows linearly with $t$ when $i$ and $j$ belong to different physical blocks. Suppose $i$ belongs to $S_a\cap B_b$, while $j$ belongs to $B_c$ but not to $S_a$, with $b\neq c$. There are $m_{ab}$ choices for $i$ and $n_c-m_{ac}$ choices for $j$. Counting all such pairs, the power of $t$ in $\Delta_{S_a}$ is
\begin{equation}
d_a({\mathcal P})=\sum_{b\neq c}m_{ab}(n_c-m_{ac}).\label{degreegeneral}
\end{equation}
Consequently, the total power of $t$ contributed by the factors $\prod_a\Delta_{S_a}^{K_a}$ is
\begin{equation}
D({\mathcal P})=\sum_aK_a d_a({\mathcal P}).\label{Ddegree}
\end{equation}
We want to show that this power is largest when the spectral blocks coincide with the physical blocks. For the physical block assignment $S_a=B_a$, every eigenvalue difference appearing in $\Delta_{S_a}$ connects distinct physical blocks. We therefore obtain
\begin{equation}
D_*=\sum_aK_an_a(N-n_a).\label{Dstar}
\end{equation}
Using the relations satisfied by $m_{ab}$, we can rewrite the general degree as
\begin{equation}
d_a({\mathcal P})=n_a(N-n_a)-\sum_bm_{ab}(n_b-m_{ab}).
\end{equation}
It follows immediately that
\begin{equation}
D_*-D({\mathcal P})=\sum_{a,b}K_a m_{ab}(n_b-m_{ab})\geq0.
\end{equation}
Every term on the right-hand side is nonnegative. If the exponents $K_a$ are positive, equality holds only when each $m_{ab}$ is either zero or $n_b$. In other words, equality is possible only when no physical block is split between different spectral blocks. Since the two partitions contain the same number of nonempty blocks, they must then agree up to permutations of complete blocks. This establishes that the physical block assignments give the largest powers of $t$ in the factors $\prod_a\Delta_{S_a}^{K_a}$. Every other assignment has a strictly smaller power. This is the general version of the gap that we found explicitly in the $SU(2)\times SU(3)$ and $SU(2)\times SU(2)$ examples.

To turn this observation into a detector identity, we must account for two additional effects. First, the corrected core operators can contribute additional powers of $t$. Second, when several assignments correspond to permutations of complete physical blocks, their contributions must be combined with the appropriate weights. As the example of two identical fermionic cores demonstrated, these weights are essential to avoid cancellations.

The first issue is straightforward. Let $A_a=A(c_a)$ denote the largest number of occurrences of the scalar letter $A^1$ in any monomial of $c_a$, and set
\begin{equation}
A_{\rm tot}=\sum_a A_a.
\end{equation}
The argument of the examples above apply without change: inverse commutators and their correction terms are $O(1)$ under the block-diagonal restriction, while the effective scalar $A^1$ grows at most linearly with $t$. Thus the product of core operators grows at most as $t^{A_{\rm tot}}$.

For the second issue, recall that our general multicore operator is
\begin{equation}
{\mathcal O}=\sum_{\mathcal P}W_{\mathcal P}\prod_a\Delta_{S_a}^{K_a}
c_a(\widehat{\mathcal L}_{S_a}).
\end{equation}
The weights $W_{\mathcal P}$ are polynomials in the eigenvalues and thus also become polynomials in $t$ under restriction. They ensure that the contributions from equivalent physical block assignments combine appropriately. We will assume that the leading contributions from these assignments do not cancel and that their sum takes the form of a nonzero constant $\Gamma$ multiplying the product of the original cores.

For simplicity, take all the exponents equal, $K_a=K$. The difference between the power of $t$ in the dressing factors for the physical block assignments and any assignment that splits a physical block is then at least $Kg_{\min}$, where
\begin{equation}
g_{\min}=\min_{{\mathcal P}\text{ outside physical block orbit}}\,\,
\sum_{a,b}m_{ab}(n_b-m_{ab})>0.
\end{equation}
The positivity of $g_{\min}$ follows from the argument above.

Let $w_*$ denote the degree in $t$ of the weights on the physical block assignments, and let $w_{\max}$ be the largest degree of the weights on all other assignments. The physical block contributions appear at degree
\begin{equation}
L_*=D_*+w_*,
\end{equation}
while the unwanted contributions have degree at most
\begin{equation}
D_*-Kg_{\min}+w_{\max}+A_{\rm tot}.
\end{equation}
Consequently, choosing $K$ sufficiently large that
\begin{equation}
Kg_{\min}>A_{\rm tot}+w_{\max}-w_*
\end{equation}
ensures that none of the unwanted assignments contributes at order $t^{L_*}$. We therefore obtain the general detector identity
\begin{equation}
[t^{L_*}]\rho_t({\mathcal O})=\Gamma\bigotimes_{a=1}^k c_a,\label{generalidentity}
\end{equation}
provided the weights have been chosen so that the physical block contributions combine into the stated nonzero coefficient. The K\"unneth theorem then gives
\begin{equation}
\bigotimes_{a=1}^k[c_a]\neq0,
\end{equation}
because every constituent class is nonzero. Since coefficient extraction after restriction commutes with $Q$, the identity \eqref{generalidentity} proves that $[{\mathcal O}]\neq0$ in the rank-$N$ BMN cohomology.

We have thus shown that the non-exactness of the constituent lower-rank operators can be used to establish the non-exactness of the multicore operator. The only additional requirement is that the weights give a nonvanishing leading contribution from the physical block assignments. The separation in powers of $t$ needed to isolate this contribution is guaranteed by the combinatorics of the partitions, provided the dressing exponent is sufficiently large.

Finally, we turn to the question of whether the operators constructed above represent fortuitous cohomology classes. To show fortuity we must show that their cohomology classes have no ancestors in the large-$N$ theory. We establish this using the same restriction map that we used to prove non-exactness. The basic idea is simple. If a cohomology class in the rank-$N$ theory descends from a class that exists at large $N$, its image under the restriction map must also descend from large-$N$ cohomology classes of the lower-rank theories. If the detector recovers a product of classes that does not belong to this stable sector, the original operator cannot have a large-$N$ ancestor. We now make this argument precise.

Let $\mathcal S$ denote the stable trace algebra generated by single traces before imposing finite-$N$ trace relations. Evaluation on traceless $N\times N$ matrices defines a map
\begin{equation}
\mathrm{ev}_N:\mathcal S\longrightarrow\mathcal A_N,
\end{equation}
which imposes the trace relations appropriate to rank $N$. Since evaluation commutes with $Q$, it also defines a map between the corresponding cohomologies. The stable sector of the rank-$N$ theory is the image of this map:
\begin{equation}
V_N=\operatorname{im}\left(H_Q(\mathcal S)\xrightarrow{\ \mathrm{ev}_{N,*}\ } H_Q(\mathcal A_N)
\right).\label{stabledef}
\end{equation}
A nonzero cohomology class outside $V_N$ is fortuitous.

To understand how the detector acts on this sector, extend the restriction map to the stable trace algebra itself. For a partition $N=n_1+\cdots+n_k$, define
\begin{equation}
\widehat\rho_t:\mathcal S\longrightarrow(\mathcal S_1\otimes\cdots\otimes\mathcal S_k)[t],\label{formalrest}
\end{equation}
where $\mathcal S_a$ is the stable trace algebra associated with the $a$th block. The map acts by substituting the block-diagonal matrices used in our restriction into each trace word. Specifically, we replace $H$ in block $a$ by $H_a+\alpha_atI_{n_a}$, while every other BMN letter is replaced by its corresponding block letter. The trace of the resulting block-diagonal expression is then written as a sum of traces over the individual blocks. For example, for the $SU(2)\times SU(3)$ restriction with shifts $\alpha_1=3$ and $\alpha_2=-2$, we have
\begin{equation}
\widehat\rho_t\bigl(T(H\phi^2\psi_2)\bigr)=T_2(H_2\phi^2_2\psi_{2,2})
+T_3(H_3\phi^2_3\psi_{2,3})+3t\,T_2(\phi^2_2\psi_{2,2})
-2t\,T_3(\phi^2_3\psi_{2,3}).\label{formalexample}
\end{equation}
Traces on the right-hand side belong to the stable algebra. This identification is allowed because we have not imposed any finite-rank trace relations.

As discussed above, the restriction commutes with $Q$. Moreover, restriction before or after imposing the finite-$N$ trace relations gives the same result. Thus,
\begin{equation}
Q\widehat\rho_t=\widehat\rho_t Q,\qquad\rho_t\circ\mathrm{ev}_N
=\left(\mathrm{ev}_{n_1}\otimes\cdots\otimes\mathrm{ev}_{n_k}\right)
\circ\widehat\rho_t.\label{commutesquare}
\end{equation}
This identity is the key to the argument. It shows that the restriction of a stable class in the rank-$N$ theory is obtained by first restricting its large-$N$ ancestor and then evaluating the resulting formal expressions in the lower-rank theories. Since the restriction commutes with $Q$, the resulting expressions are again representatives of stable cohomology classes.

In particular, recall that our detector $\beta_h$ extracts the coefficient of $t^h$ after restriction. Equation~\eqref{commutesquare} implies
\begin{equation}
\beta_{h,*}(V_N)\subseteq V_{\mathrm{target}},\label{stablepreserve}
\end{equation}
where $V_{\mathrm{target}}$ is the image of the cohomology of the tensor product of the stable algebras under evaluation in the lower-rank theories. By the K\"unneth theorem, this stable sector is
\begin{equation}
V_{\mathrm{target}}=V_{n_1}\otimes\cdots\otimes V_{n_k}.
\end{equation}
Thus, the detector maps stable classes of the original theory into the stable sector of the product theory.

We can now complete the argument. We have already shown that, for suitable choices of the dressing exponents and weights, the detector recovers
\begin{equation}
\beta_{L_*}(\mathcal O)=\Gamma\,c_1\otimes\cdots\otimes c_k,\qquad \Gamma\neq0.
\end{equation}
Suppose that the original operator represents a stable cohomology class, $[\mathcal O]\in V_N$. Equation~\eqref{stablepreserve} then implies that
\begin{equation}
\Gamma\,[c_1]\otimes\cdots\otimes[c_k]\in V_{\mathrm{target}}.
\end{equation}
But if the constituent classes are nonzero and fortuitous, their tensor product is nonzero and outside the tensor product of the stable sectors. This is a contradiction. Therefore,
\begin{equation}
[\mathcal O]\notin V_N.
\end{equation}
Since we have already established that $[\mathcal O]\neq0$, we conclude that the multicore operator represents a fortuitous cohomology class. Thus known lower-rank fortuitous classes can be combined to construct new fortuitous classes at higher rank. The construction produces gauge-invariant, $Q$-closed polynomial operators, while the restriction map establishes both their non-exactness and the absence of large-$N$ ancestors. In this way, the cohomological properties of the original building blocks provide a direct way to establish the fortuity of the resulting multicore operators.

\section{Example: two cores \texorpdfstring{$O_{24}$}{O24} in \texorpdfstring{$SU(4)$}{SU(4)}}
\label{sec:ex22}

In this section we illustrate the general construction by combining two copies of the known $SU(2)$ fortuitous operator $O_{24}$ to obtain a new fortuitous operator in the $SU(4)$ theory. This example is interesting because both building blocks are nonabelian, so the resulting operator has two genuine cores rather than one core accompanied by spectators \cite{DKL}. We will construct the operator explicitly, determine a sufficient power of $\Delta_S$ to remove its denominators, and show how the remaining eigenvalue dependence is eliminated. We also establish an exact restriction identity that recovers the product of the two original $SU(2)$ classes. Unlike the general construction, this identity holds without requiring a large separation between the powers of $t$ contributed by different partitions.

We use the $SU(2)$ fortuitous threshold representative \cite{ChangLin22,CKLP,CKLLP},
\begin{equation}
\begin{gathered}
O_{24}=\epsilon^{p_1p_2p_3}\, v^m{}_{p_2}\,v^n{}_{p_3}\,
\tr\big(\psi_m\psi_n\psi_{p_1}\big),\\
v^m{}_n=\tr(\phi^m\psi_n)-\frac13\delta^m_n\tr(\phi^p\psi_p).
\end{gathered}\label{eq:22core}
\end{equation}
This operator is fermionic and represents a nontrivial fortuitous cohomology class in the $SU(2)$ theory. Let $H=\phi^1$ have eigenvalues $\lambda_1,\ldots,\lambda_4$, and divide the four colours into two pairs $S$ and $T$. Using the spectral projectors and corrected effective letters constructed in Sections~\ref{effletersec} and \ref{ginvsec}, we can evaluate $O_{24}$ on each pair. Denote the resulting operators by
\begin{equation}
C_S=O_{24}(\widehat{\mathcal L}_S),\qquad C_T=O_{24}(\widehat{\mathcal L}_T).\label{eq:22CS}
\end{equation}
Since $O_{24}$ is fermionic, exchanging the two cores changes the sign of their product. As explained in Section~\ref{DetcorSec}, we compensate for this sign by introducing the antisymmetric weight
\begin{equation}
w_{S,T}=\frac12\sum_{i\in S}\lambda_i-\frac12\sum_{j\in T}\lambda_j.\label{eq:22weight}
\end{equation}
Under exchange of the two blocks, $w_{T,S}=-w_{S,T}$, so the weight compensates for the sign change of the fermionic operators. Our multicore operator is therefore
\begin{equation}
O^{(K)}_{2,2}=\sum_{(S,T)}w_{S,T}\Delta_S^K\Delta_T^K C_SC_T
=2\sum_{\{S,T\}}w_{S,T}\Delta_S^{2K}C_SC_T.\label{eq:22op}
\end{equation}
The second expression follows because $\Delta_T=\Delta_S$ and the two orderings of each partition contribute equally. The resulting operator is $Q$-closed. Our next task is to determine a sufficient value of $K$ to make it polynomial.

We first give the explicit projector formulas needed for this example. Consider $S=\{1,2\}$ and $T=\{3,4\}$; the other splittings follow by relabelling the eigenvalues. Since $H$ is traceless, the weight becomes
\begin{equation}
w_{S,T}=\lambda_1+\lambda_2=-(\lambda_3+\lambda_4).
\end{equation}
For simplicity, write $w=w_{S,T}$. The polynomials introduced in Section~\ref{ginvsec} are
\begin{equation}
q_S(x)=x^2-wx+\lambda_1\lambda_2,\qquad r_S(x)=x^2+wx+\lambda_3\lambda_4.
\end{equation}
The Bézout identity is solved by
\begin{equation}
\begin{aligned}
a_S(x)&=2w^2+\lambda_1\lambda_2-\lambda_3\lambda_4+2wx,\\
b_S(x)&=2w^2-\lambda_1\lambda_2+\lambda_3\lambda_4-2wx.
\end{aligned}\label{eq:22ab}
\end{equation}
Consequently, the projectors can be written as
\begin{equation}
P_S=\frac{b_S(H)r_S(H)}{\Delta_S},\qquad P_T=\frac{a_S(H)q_S(H)}{\Delta_S}.
\end{equation}
The inverse commutators take the explicit form
\begin{equation}
\begin{aligned}
X_S(Y)&=\frac{b_S(H)}{\Delta_S}P_S(HY+YH+wY)P_T,\\
X_T(Y)&=\frac{a_S(H)}{\Delta_S}P_T(HY+YH-wY)P_S.
\end{aligned}\label{eq:22X}
\end{equation}
These formulas allow us to construct the corrected effective letters directly from the original $4\times4$ matrices. In particular, with $X_m=X_S(\psi_m)$ and
\begin{equation}
Z^0=Z-\frac12\tr(Z)P_S,
\end{equation}
the effective letters required for $O_{24}$ are
\begin{equation}
\begin{aligned}
\widehat A^m&=(P_S\phi^mP_S)^0,\qquad
\widehat\psi_{2,3}=(P_S\psi_{2,3}P_S)^0,\\
\widehat\psi_1&=
(P_S\psi_1P_S-X_3\phi^3P_S-X_2\phi^2P_S)^0.
\end{aligned}\label{eq:22letters}
\end{equation}
The letters associated with $T$ are obtained by exchanging $S$ and $T$. All traces are ordinary traces of the original $4\times4$ matrices.

We can now determine the power of $\Delta_S$ needed to remove the denominators. The general argument of Section~\ref{ginvsec} guarantees polynomiality for $K\geq35$, since $O_{24}$ contains seven letters. For this core, however, we can obtain a substantially better bound.

The improvement comes from the structure of the traces. Neighbouring projectors inside a trace combine according to $P_S^2=P_S$, reducing the number of denominators. Each effective letter therefore requires at most one inverse power of $\Delta_S$ from the projectors. The corrected fermion $\widehat\psi_1$ can require two additional inverse powers, arising from the inverse commutator and its complementary projector. Taking the traceless part does not increase the count, since the additional projector can combine with those already in the trace.

The resulting trace-level denominator estimate is $\ell+2n_{\psi_1}$ for a monomial with $\ell$ effective letters, of which $n_{\psi_1}$ are corrected fermions of type $\widehat\psi_1$; this is the refined counting used in \cite{DKL}. It is important that the estimate is applied to the complete trace monomial, where adjacent projectors cancel, rather than separately to individual matrix entries. Every term in $O_{24}$ contains seven letters and at most three occurrences of $\psi_1$. Consequently,
\begin{equation}
\Delta_S^{13}C_S
\end{equation}
is polynomial in the original BMN letters and the eigenvalues of $H$. The same argument applies to $C_T$. We may therefore take
\begin{equation}
K\in\mathbb Z,\qquad K\geq13.\label{eq:22K}
\end{equation}
This bound is considerably smaller than the general sufficient bound and ensures that every summand in \eqref{eq:22op} is polynomial in the original matrix entries and the eigenvalues.

It is useful to record the charges of the resulting operator. Let $R_m$ be the three Cartan charges of $SO(6)_R$, and let $J_1,J_2$ denote the two angular momenta. Using
\begin{equation}
j=2(R_1+R_2+R_3)+3(J_1+J_2),
\end{equation}
we find $R_1=4+8K$, $R_2=R_3=3$, $J_1+J_2=10$, $j=50+16K$, and $\ell=15+8K$, where $\ell$ is the total number of letters.

We have established that the denominators can be removed, but the resulting expressions still depend explicitly on the eigenvalues of $H$. We now show more carefully how this dependence can be eliminated. The key simplification is that there are only three unordered ways of dividing four colours into two pairs. For a fixed splitting $S=\{1,2\}$ and $T=\{3,4\}$, write
\begin{equation}
 w=\lambda_1+\lambda_2=-(\lambda_3+\lambda_4),\qquad
 A=\lambda_1\lambda_2,\qquad B=\lambda_3\lambda_4.
\end{equation}
The characteristic polynomial of the traceless $4\times4$ matrix $H$ is
\begin{equation}
\det(zI-H)=z^4+e_2z^2-e_3z+e_4,\label{eq:22char}
\end{equation}
where, writing $p_k=\tr(H^k)$,
\begin{equation}
e_2=-\frac12p_2,\qquad e_3=\frac13p_3,\qquad e_4=\frac18p_2^2-\frac14p_4.
\end{equation}
There is a subtlety in applying symmetric-polynomial methods: in an eigenbasis, eigenvalue indices are tied to matrix indices. As explained in Section~\ref{ginvsec}, this dependence must first be encoded by polynomial matrix operations, for example $(\lambda_i-\lambda_j)Y_{ij}=[H,Y]_{ij}$. After canceling denominators and doing so, we may expand the expression in fixed matrix words and traces. The remaining labelled eigenvalues appear only in scalar polynomial coefficients. It is these coefficients, rather than the matrix indices, that we now eliminate. For a fixed splitting, permutations within either pair leave the expression unchanged. The polynomial coefficients can therefore be written in terms of $w,A,B$. Exchanging the pairs acts as
\begin{equation}
(w,A,B)\longmapsto(-w,B,A).
\end{equation}
In the full summand, this exchange changes the sign of both the antisymmetric weight and the product of the two fermionic cores, so the result is unchanged. The invariant polynomials in $w,A,B$ are generated by
\begin{equation}
 w^2,\qquad A+B,\qquad AB,\qquad w(B-A).
\end{equation}
For our traceless $H$, these combinations are
\begin{equation}
 w^2=x_{\mathcal P},\qquad A+B=e_2+x_{\mathcal P},\qquad
 AB=e_4,\qquad w(B-A)=e_3,\label{eq:22inv}
\end{equation}
where $x_{\mathcal P}=w_{S,T}^2$ labels the unordered splitting $\mathcal P=\{S,T\}$. Thus, once the matrix-index dependence has been incorporated into matrix multiplication, the remaining coefficients can be expressed in terms of $x_{\mathcal P}$ and $e_2,e_3,e_4$. These quantities satisfy a cubic relation. Indeed,
\begin{equation}
 e_3^2=w^2(B-A)^2 =x_{\mathcal P}\big[(e_2+x_{\mathcal P})^2-4e_4\big].
\end{equation}
The three values of $x_{\mathcal P}$ are therefore the roots of
\begin{equation}
 \chi(x)=x^3+2e_2x^2+(e_2^2-4e_4)x-e_3^2,\qquad \chi(x_{\mathcal P})=0.\label{eq:22cubic}
\end{equation}
A useful additional identity is $\Delta_S=\chi'(x_{\mathcal P})$. For each unordered splitting, put
\begin{equation}
 T_{\mathcal P}^{(K)}=w_{S,T}\Delta_S^{2K}C_SC_T,
 \qquad O_{2,2}^{(K)}=2\sum_{\mathcal P}T_{\mathcal P}^{(K)}.
\end{equation}
The invariant-coefficient argument above gives a common polynomial expression for $T_{\mathcal P}^{(K)}$ in the variable $x_{\mathcal P}$, with coefficients that are polynomial matrix expressions independent of the chosen splitting. Since $\chi$ is monic, division by $\chi(x)$ produces a remainder of degree at most two
\begin{equation}
 T_{\mathcal P}^{(K)}=\tau_0^{(K)}+\tau_1^{(K)}x_{\mathcal P}
 +\tau_2^{(K)}x_{\mathcal P}^2.\label{eq:22remainder}
\end{equation}
The $\tau_i^{(K)}$ are polynomial expressions in the original BMN matrices and the characteristic coefficients of $H$; they are the same for all three splittings. This is an algebraic reduction, not an explicit expansion of these potentially lengthy multi-trace coefficients. The roots of $\chi$ obey
\begin{equation}
 \sum_{\mathcal P}x_{\mathcal P}=p_2,\qquad
 \sum_{\mathcal P}x_{\mathcal P}^2=\frac32p_2^2-2p_4.
\end{equation}
Summing \eqref{eq:22remainder} over the three splittings therefore gives
\begin{equation}\label{eq:22elim}
 O_{2,2}^{(K)}=6\tau_0^{(K)}+2p_2\tau_1^{(K)}
 +(3p_2^2-4p_4)\tau_2^{(K)}.
\end{equation}
Every quantity on the right is polynomial in the original BMN letters. This establishes the elimination of the individually labelled eigenvalues and gives a finite reduction procedure, without requiring the fully expanded multi-trace expression.

We now establish a particularly simple restriction identity for this operator. As in Section \ref{DetcorSec}, divide the four colours into two physical blocks $B_1$ and $B_2$ of size two and take
\begin{equation}
\rho_t(H)=(H_1+2tI_2)\oplus(H_2-2tI_2),
\end{equation}
with all other BMN letters block diagonal. Let $\mu_1,\mu_2$ and $\nu_1,\nu_2$ denote the eigenvalues of $H_1$ and $H_2$, respectively. When the selected spectral blocks coincide with the physical blocks, the corrected effective letters reduce to the original $SU(2)$ letters. The corresponding terms therefore recover the product of the two $O_{24}$ operators, multiplied by the appropriate powers of the eigenvalue differences. What happens to the remaining partitions? Every mixed partition selects one colour from each physical block. Since the restricted matrices are block diagonal, the projected effective letters are diagonal in the selected two-dimensional subspace. In particular, the traceless effective fermions are proportional to the same diagonal matrix ${\rm diag}(1,-1)$. The trace of a product of three such matrices vanishes. Since every term in $O_{24}$ contains a trace of three fermions, the entire core operator vanishes on these mixed partitions. Thus, the unwanted partitions vanish identically, rather than merely contributing lower powers of $t$. Only the two physical block assignments survive, and we obtain the exact identity
\begin{equation}
\rho_t\bigl(O_{2,2}^{(K)}\bigr)={}8t\prod_{i=1}^{2}\prod_{j=1}^{2}(4t+\mu_i-\nu_j)^{2K} O_{24}(\mathcal L_1)O_{24}(\mathcal L_2).\label{eq:22restr}
\end{equation}
Extracting the coefficient of $t^{8K+1}$ gives
\begin{equation}
[t^{8K+1}]\rho_t\bigl(O_{2,2}^{(K)}\bigr)=8\cdot4^{8K}\,
O_{24}(\mathcal L_1)O_{24}(\mathcal L_2).\label{eq:22detector}
\end{equation}

Unlike the general detector argument, this result does not require an additional condition on $K$ to suppress unwanted partitions. The only restriction is the polynomiality bound $K\geq13$. Since the two copies of $O_{24}$ represent nonzero fortuitous classes, the arguments of Section~\ref{DetcorSec} establish that the resulting $SU(4)$ operator is also non-exact and fortuitous. The product of the two cores is nonzero even though $O_{24}$ is fermionic and hence $O_{24}^2=0$ within a single $SU(2)$ theory. The two copies belong to different physical blocks, so their product represents a nonzero class in the tensor product of the two $SU(2)$ cohomologies.

This example illustrates several features of the multicore construction. First, it produces a fortuitous operator from two independent nonabelian cores, without introducing spectator colours. Second, the detailed structure of $O_{24}$ allows us to improve the general polynomiality bound substantially. Finally, the restriction is particularly simple: the mixed partitions vanish identically, so the original product of cohomology classes is recovered through an exact identity, without requiring the additional degree-gap condition of the general construction. The example also illustrates why it is important to allow restrictions with several nonabelian blocks. A restriction retaining only a single nonabelian core cannot detect the two-core product in this example. The multicore detector, by contrast, recovers the nontrivial cohomology associated with both cores simultaneously.

\section{Discussion}\label{sec:discussion}

In this paper, building on the construction of~\cite{DKL}, we have developed a systematic method for constructing fortuitous operators in the $SU(N)$ BMN theory from lower-rank building blocks. For any partition $N=n_1+\cdots+n_k$, we construct effective BMN letters associated with the individual blocks, each obeying the supersymmetry transformations of the corresponding lower-rank theory. Known $Q$-closed operators can then be evaluated on these effective letters and combined to produce gauge-invariant, $Q$-closed polynomial operators of the full $SU(N)$ theory. The essential additional result is that a suitable restriction map allows us to recover the original lower-rank cohomology classes. This establishes non-exactness without requiring a brute-force computation of the higher-rank cohomology. By tracking the action of the restriction map on stable classes, we can also establish fortuity whenever the detected product lies outside the stable sector.

The construction provides a useful picture of how fortuitous cohomology is organized at finite $N$. The original $N\times N$ matrices contain several sets of effective lower-rank letters, each associated with a different block of colour indices. Although these effective letters are constructed from the same original matrices and need not be algebraically independent, each set obeys the appropriate supersymmetry algebra. The corrections to the projected letters account for their coupling to the remaining colours. In the separated-eigenvalue picture, the different blocks can be viewed as distinct stacks of colours, while the factors of eigenvalue differences used to remove denominators are naturally associated with the off-diagonal degrees of freedom connecting these stacks. This provides an intuitive interpretation of the construction. Its validity is established through algebraic identities proved in the paper.

An important consequence is that fortuitous cohomology admits a form of composition. A fortuitous class need not arise from a single nonabelian core: several lower-rank classes can contribute simultaneously to a new class of the larger theory. The $SU(4)$ example provides an explicit realization of this possibility, combining two copies of the $SU(2)$ fortuitous operator $O_{24}$. The resulting operator represents a nontrivial fortuitous cohomology class, even though the two constituent operators are fermionic and the square of either vanishes within a single $SU(2)$ theory. The restriction map distinguishes the two cores and recovers their nonzero tensor product. This suggests that the multiplicative structure of finite-rank cohomology, and the distinction between composite and genuinely new classes, deserve further investigation.

The possibility of constructing large families of fortuitous operators is particularly interesting in the context of black hole microstates. Fortuitous classes provide natural candidates for the enormous number of supersymmetric states required to account for the entropy of $1/16$-BPS black holes. Our construction shows how known lower-rank classes generate higher-rank families, but it does not establish that these families have the multiplicity or charge distribution needed to reproduce black hole entropy. Indeed, operators constructed from a fixed number of small cores dressed by eigenvalue differences appear more naturally associated with brane-like configurations than with macroscopic black holes. This raises an important question: do black hole microstates require fortuitous cores whose rank grows with $N$, or can sufficiently complicated combinations of smaller cores already produce the required number of states? Understanding the cohomology associated with genuinely large cores, and whether such cores can be related to combinations of smaller ones, is an important direction for future work.

A further question is whether these constructions survive quantum corrections. Our analysis concerns the classical BMN cohomology, which describes the one-loop BPS spectrum. Quantum corrections can lift classical cohomology classes, so their nontriviality at this order does not by itself establish the existence of exact BPS states \cite{ChoiKim,KLLS26,KKLP26}. It would be interesting to determine whether the effective-letter construction and restriction maps can be extended to the quantum-corrected supercharge. Such an extension could provide a systematic way of constructing protected states with several nonabelian cores.

The main lesson of our analysis is that fortuitous cohomology has more structure than is apparent from the study of individual operators at small rank. Lower-rank fortuitous classes can be combined, their non-exactness can be detected without solving the full higher-rank cohomology, and several nonabelian cores can coexist within a single gauge-invariant operator. Understanding how far this compositional structure extends, and what additional structures are needed to account for black hole microstates, now becomes a concrete problem.

\acknowledgments
RdMK would like to thank Kyung Hee University and KIAS for their hospitality and support, during which this work was completed. 
The work was supported by a start-up research fund of Huzhou Normal University (RdMK, MK), a Zhejiang Province talent award (RdMK), a Changjiang Scholar award (RdMK), the Shuimu Tsinghua Scholar Program of Tsinghua University (SD), the Beijing Natural Science Foundation of China Grant No. IS25035 (SD),
the National Research Foundation of Korea (NRF) grant funded by the Korean government (MSIT) (RS-2025-25414114 (MK), RS-2022-NR069038 (SL)), and the Brain Pool Program funded by the Ministry of Science and ICT through the National Research Foundation of Korea (RS-2025-25457100) (RdMK,SL).

\appendix

\section{An example involving the field-strength letter}\label{Appflett}

The $SU(4)$ example studied in Section~\ref{sec:ex22} combines two copies of the fortuitous operator $O_{24}$, which depends only on the scalar and fermionic BMN letters. It is useful to also consider an example involving the auxiliary field-strength letter $f$, since this tests the corrections to the effective letters developed in Section~\ref{effletersec}.

For this purpose, we use the known $SU(2)$ fortuitous operator $O_{36}$ of \cite{CKLLP}, which is the next member of the tower containing $O_{24}$. An explicit representative is
\begin{equation}
\begin{split}
O_{36}={}&\tr(ff)\,O_{24}\\
&+\epsilon^{n_1n_2n_3}\epsilon^{p_1p_2p_3}\tr(f\psi_{n_1})\tr(\phi^m\psi_{p_1})
\tr(\psi_{n_2}\psi_{p_2})\tr(\psi_m\psi_{n_3}\psi_{p_3})\\
&+\frac1{36}\epsilon^{m_1m_2m_3}\epsilon^{n_1n_2n_3}\epsilon^{p_1p_2p_3}
\tr(\psi_{m_1}\psi_{n_1}\psi_{p_1})\tr(\psi_{m_2}\psi_{n_2}\psi_{p_2})
\tr(\psi_{m_3}\psi_{n_3}\psi_{p_3}).
\end{split}
\end{equation}
We embed this operator into the $SU(3)$ theory by selecting a pair of colours $S$, with the remaining colour acting as a spectator. Using the corrected effective letters $\widehat{\mathcal L}_S$ constructed in Section~\ref{effletersec}, define
\begin{equation}
\mathcal O_{2,1}^{(K)}=\sum_{|S|=2}\Delta_S^K O_{36}(\widehat{\mathcal L}_S).
\end{equation}
The sum runs over the three possible choices of the spectator colour. As established in Section~\ref{ginvsec}, multiplying by a sufficiently high power of $\Delta_S$ removes the denominators, and summing over the possible choices of $S$ eliminates the dependence on individual eigenvalues. The resulting operator is a gauge-invariant, $Q$-closed polynomial in the original $SU(3)$ letters.

For this particular representative, the refined denominator counting used in Section~\ref{sec:ex22} gives the sufficient bound $K\geq19$. Indeed, every term of $O_{36}$ contains nine letters, with at most three occurrences of $\psi_1$ and two occurrences of $f$. Each corrected $\psi_1$ or $f$ requires at most two additional powers of $\Delta_S$, giving the bound
\begin{equation}
9+2(3)+2(2)=19.
\end{equation}

To establish non-exactness and fortuity, restrict the matrices to an $SU(2)$ block and a spectator
\begin{equation}
\rho_t(H)=(H_2+tI_2)\oplus(-2t),\qquad\rho_t(Y)=Y_2\oplus0,\quad Y\neq H.
\end{equation}
When the selected pair coincides with the physical $SU(2)$ block, the effective letters reduce to the original $SU(2)$ letters. Writing the eigenvalues of $H_2$ as $\mu_1,\mu_2$, the corresponding dressing factor becomes
\begin{equation}
\rho_t(\Delta_S)=(3t+\mu_1)(3t+\mu_2)=9t^2+\mu_1\mu_2.
\end{equation}

For either of the other two choices of $S$, the projected fermions and their corrections are diagonal and traceless, so they are proportional to $\operatorname{diag}(1,-1)$. Their triple traces therefore vanish. Since every term of $O_{36}$ contains such a triple trace, both unwanted contributions vanish identically. Consequently, we obtain the exact restriction identity
\begin{equation}
\rho_t(\mathcal O_{2,1}^{(K)})=(9t^2+\mu_1\mu_2)^K O_{36}(\mathcal L_2).
\end{equation}
Extracting the coefficient of $t^{2K}$ gives
\begin{equation}
\beta_{2K}(\mathcal O_{2,1}^{(K)})=9^K O_{36}(\mathcal L_2).
\end{equation}
Since $O_{36}$ represents a nonzero fortuitous class in the $SU(2)$ theory, the restriction argument of Section~4 establishes that $\mathcal O_{2,1}^{(K)}$ is also non-exact and fortuitous for $K\geq19$.

This example confirms that the construction works for operators involving the field-strength letter $f$. It also provides another instance in which the unwanted partitions vanish exactly, so that the fortuitous building block is recovered without requiring an additional degree-gap condition.

\end{document}